\documentclass[letterpaper,10pt]{article}

\usepackage[dvips]{graphicx}
\usepackage[]{caption}
\usepackage{morefloats}
\usepackage{color}
\usepackage[rgb,dvipsnames]{xcolor}
\definecolor{grey}{gray}{0.7}
\usepackage{amsmath}
\usepackage{amssymb}
\usepackage{latexsym}
\usepackage{enumitem}
\usepackage[square,authoryear]{natbib}
\usepackage[compact]{titlesec}
\usepackage{fancyhdr}
\usepackage[textwidth=6.5in, textheight=8.5in, top=1.0in, bottom=1.0in, left=0.875in, right=0.875in]{geometry}
\usepackage{setspace}
\usepackage{indentfirst}
\usepackage{lastpage}     %%  Allow user to know total # of pages (e.g. page 5 of 9)
\setlist{leftmargin=7mm}
\titleformat{\section}[hang]{\Large\bfseries}{\thesection}{10pt}{}  %% horizontal spacing
\titleformat{\subsection}[hang]{\large\bfseries}{\thesubsection}{7pt}{}
\titleformat{\subsubsection}[hang]{\normalsize\bfseries}{\thesubsubsection}{5pt}{}
\newcommand{\myname}{Lynn B. Wilson III}
\newcommand{\papertitle}{On the role of wave-particle interactions in the macroscopic dynamics of collisionless plasmas}
\newcommand{\shorttitle}{WAVE-PARTICLE INTERACTIONS}
\newcommand{\headertext}{L.B. WILSON III ET. AL.}
\newcommand{\mpageofn}{\thepage\ of \pageref{LastPage}}
\newcommand{\evenhdrnote}{\shorttitle}
\newcommand{\oddhdrnote}{\headertext}
\newcommand{\footnoteremember}[2]{\footnote{#2}\newcounter{#1}\setcounter{#1}{\value{footnote}}}

\begin{document}
%%  style of bibliography
%\bibliographystyle{agu08}

%%  TITLE
\title{\bf \papertitle}
%%  AUTHORS AND AFFILIATIONS - 2 methods
\author{\myname\footnoteremember{1}{NASA Goddard Space Flight Center, Greenbelt, Maryland, USA.},
A.W. Breneman\footnoteremember{2}{School of Physics and Astronomy, University of Minnesota, Minneapolis, Minnesota, USA.},
A. Osmane\footnoteremember{3}{Department of Radio Science, Aalto University, Finland, 02150.},
D.M. Malaspina\footnoteremember{4}{University of Colorado at Boulder, LASP, Colorado, USA.},
}
\date{\today}
\maketitle

%%  ABSTRACT 
\begin{abstract}
  Here we present a commentary on the role of small-scale (e.g., few tens of meters) wave-particle interactions in large-scale (e.g., few tens of kilometers) processes and their capacity to accelerate particles from thermal to suprathermal or even to cosmic-ray energies.  The purpose of this commentary is to provoke thought and further investigation into the relative importance of electromagnetic waves in the global dynamics of many macroscopic systems.  We specifically focus on the terrestrial radiation belts and collisionless shock waves as examples.
\end{abstract}

%%----------------------------------------------------------------------------------------
%%  Section:  Background and Motivation
%%----------------------------------------------------------------------------------------
\section{Background and Motivation}  \label{sec:introduction}
\indent  What is the relative importance of small-scale/microphysical (i.e., electron gyroradii to electron Debye scales) plasma processes to the acceleration of particles from thermal to high energies?  Additionally, can these microphysical plasma processes influence or even dominate macroscopic (i.e., greater than ion gyroradii to system size) processes, thus affecting global dynamics?  These are fundamental and unresolved questions in plasma \citep[e.g.,][]{balogh13a, cattell12a} and astrophysical research \citep[e.g.,][]{bykov14a}.

\indent  Recent observations of large amplitude electromagnetic waves in the terrestrial radiation belts \citep[i.e.,][]{cattell08a, kellogg10c, wilsoniii11e} and in collisionless shock waves \citep[i.e.,][]{wilsoniii14a, wilsoniii14b} have raised questions regarding the macrophysical effect of these microscopic waves on these two systems.  These observations have renewed and increased attention on the relative importance of wave-particle interactions in these regions of space.  In this commentary, we provide a few examples that illustrate the potential of small-scale electromagnetic waves to quickly and strongly energize particles and affect the global dynamics of macroscopic systems.  We then provide a means to extrapolate impact of wave-particle interactions to regions of space that are currently inaccessible (e.g., magnetar atmospheres).

%%----------------------------------------------------------------------------------------
%%  Section:  Radiation Belt Dynamics
%%----------------------------------------------------------------------------------------
\section{Radiation Belt Dynamics}  \label{sec:RadiationBeltDynamics}
\indent  The discovery of large amplitude whistler mode waves (LAWs) \citep[i.e.,][]{cattell08a} fundamentally changed our evaluation of the role played by small-scale wave-particle interactions in the terrestrial radiation belts.  Most prior studies of whistler mode waves (i.e., including chorus and hiss) in the inner magnetosphere, relying upon time- and frequency-averaged dynamic spectra, observed typical amplitudes of only $\sim$0.5 mV/m \citep[e.g.,][]{meredith01a} and $\sim$0.01--0.1 nT \citep[e.g.,][]{horne05a}.  In contrast, recent high resolution time series measurements have observed waves with magnetic amplitudes $\delta B$ $>$4--8 nT (peak-to-peak) \citep[e.g.,][]{wilsoniii11e, santolik14a} and electric amplitudes $\delta E$ $>$ 200--400 mV/m \citep[e.g.,][]{agapitov14a, cattell08a, kellogg10c, kellogg11a, wilsoniii11e}, where amplitudes of several 10s of mV/m and/or several hundred pT have occurrence rates reaching $<$ 10 \% \citep[e.g.,][]{cully08a, li11a}.  Below we discuss potential impacts of such large amplitude waves in order to provoke further research into their relative importance.

%%****************************************************************************************
%%  Subsection:  Nonlinear Acceleration
%%****************************************************************************************
\subsection{Nonlinear Acceleration}  \label{subsec:NonlinearAcceleration}
\indent  \citet[][]{wilsoniii11e}, using fully nonlinear equations \citep[e.g., Equation 44 in][]{omura07a}, derived a rough estimate of the maximum kinetic energy gain of an electron interacting with the 8 nT LAW shown in Figure \ref{fig:exampledWWEvsJoExtrap}A, finding an energy increase between $\sim$30--70 MeV, energies 30--70 times larger than current measurements.  The equation assumes the waves propagate parallel to the quasi-static magnetic field and that, somewhat unrealistically, the waves are constant amplitude and omnipresent along the particle trajectory.  Even so, dynamical simulations have shown that keV electrons can be accelerated to hundreds of keV energies \citep[e.g.,][]{artemyev14e, artemyev14g, osmane16a} or even MeV energies on kinetic time scales (i.e., trapping times and/or a few tens of gyroperiods, which corresponds to $\sim$0.1--10 ms in the terrestrial radiation belts) by oblique whistler mode waves \citep[e.g.,][]{kersten11b, osmane12a}.

\indent  To further illustrate the significance of including the fully nonlinear relations, we compare with a simple linear approximation where we estimate the wave electric potential by assuming $\phi$ $\sim$ $2 \ \pi \ \lvert \delta \mathbf{E} \rvert / k$ $=$ $\lvert \delta \mathbf{E} \rvert \ \lambda$.  We can use the observed wave amplitude shown in Figure \ref{fig:exampledWWEvsJoExtrap}A and previous estimates of of the wavenumber ranges for whistler modes of $k \rho{\scriptstyle_{ce}}$ $\sim$ 0.2--1.0 \citep[e.g.,][]{hull12a, wilsoniii13a} to find $\phi$ $\lesssim$ 4 keV.  Thus, one can see that the simple linear approximation is up to several orders of magnitude smaller than the fully nonlinear estimates shown above.

%%****************************************************************************************
%%  Subsection:  Global Dynamics
%%****************************************************************************************
\subsection{Global Dynamics}  \label{subsec:GlobalDynamics}
\begin{figure}[!htb]
  \centering
    {\includegraphics[trim = 0mm 0mm 0mm 0mm, clip, width=140mm]
    {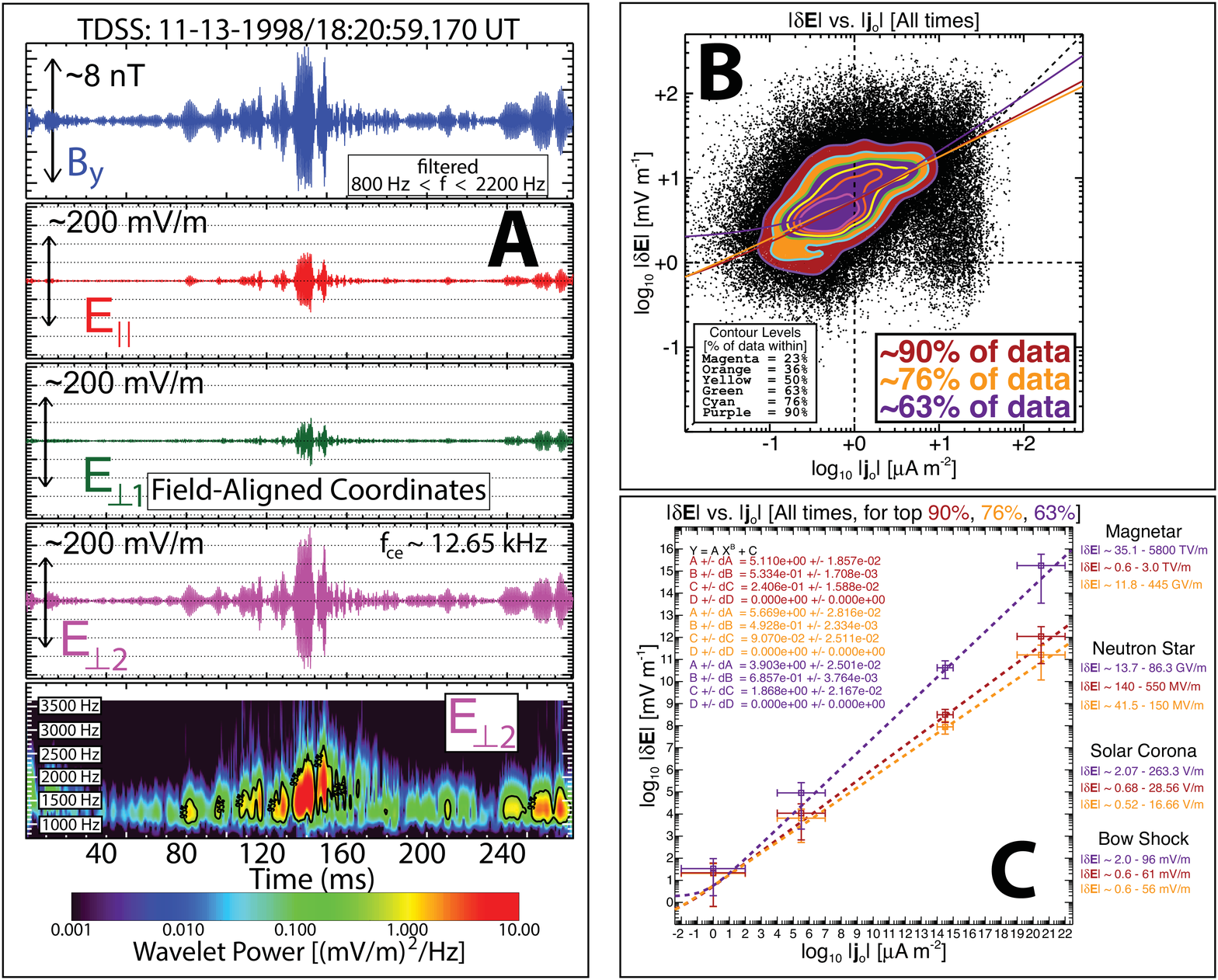}}
    \caption[Whistler, $\delta \textbf{E}$ vs. $\textbf{j}{\scriptstyle_{o}}$, and Extrapolations]{The image shows an example waveform capture of a LAW in the terrestrial radiation belts (panel A), plot of the correlation between $\left\lvert \delta \textbf{E} \right\rvert$ and $\left\lvert \textbf{j}{\scriptstyle_{o}} \right\rvert$ (panel B), and an extrapolation of the fit lines shown in panel B to different regions of space with fit parameters color-coded and labeled (panel C).  See text for more details.  Panel A is adapted from Figure 2 in \citet[][]{wilsoniii11e} and panel B is adapted from Figure II:2 in \citet[][]{wilsoniii14b}.}
    \label{fig:exampledWWEvsJoExtrap}
\end{figure}

\indent  We use the example waveform in Figure \ref{fig:exampledWWEvsJoExtrap}A to discuss global radiation belt time-scales.  The wave had a lower bound (i.e., due to saturation and missing two magnetic field components) on its in situ Poynting flux of $\sim$300 $\mu W \ m^{-2}$, which is roughly four orders of magnitude larger than the estimates found by \citet{santolik10a} (i.e., $\sim 5 \times 10^{-2} \mu W \ m^{-2}$).  They discussed the potential for chorus to accelerate plasma sheet electrons up to 1 MeV energies, assuming a typical radiation belt flux of $\sim 10^{6} \ m^{-2} \ s^{-1}$.  To produce this, they argued that the chorus waves would need to deposit an energy density of $\sim 5 \times 10^{-12} J \ m^{-3}$.  They assumed a field-aligned column $\sim$3 $R{\scriptstyle_{E}}$ in length, which corresponds to a column-integrated energy density of $\sim 10^{-4} J \ m^{-2}$.  Assuming a 1\% efficiency, they found that the chorus waves could produce such a column on a time scale of days.

\indent  If we follow the same line of reasoning, but with our observation of $\sim$300 $\mu W \ m^{-2}$, we find a time-scale of $\sim$33 seconds.  While the time-scale is much longer than the duration of the wave packet, the dramatic difference with previous estimates illustrates the potential impact of such LAWs on the global system.  This argument is supported by recent work \citep[e.g.,][]{artemyev15a} showing that oblique whistler mode waves can carry up to 80\% of the energy in resonant wave-particle interactions, significantly altering radiation belt particle lifetimes, lending support to the argument that LAWs play an important or even dominant role in the global dynamics of the terrestrial radiation belts \citep[e.g.,][]{agapitov14a, wilsoniii11e}.

\indent  Below we will move on to discuss the importance of small-scale wave-particle interactions on the dynamics of another macroscopic system, collisionless shock waves.

%%++++++++++++++++++++++++++++++++++++++++++++++++++++++++++++++++++++++++++++++++++++++++
%% Image:  Whistler, ∂E vs. Jo, and Extrapolations
%%++++++++++++++++++++++++++++++++++++++++++++++++++++++++++++++++++++++++++++++++++++++++
%%++++++++++++++++++++++++++++++++++++++++++++++++++++++++++++++++++++++++++++++++++++++++
%% Image:  Whistler, ∂E vs. Jo, and Extrapolations
%%++++++++++++++++++++++++++++++++++++++++++++++++++++++++++++++++++++++++++++++++++++++++

%%----------------------------------------------------------------------------------------
%%  Section:  Collisionless Shock Wave Dynamics
%%----------------------------------------------------------------------------------------
\section{Collisionless Shock Wave Dynamics}  \label{sec:CollisionlessShockWaveDynamics}
\indent  Early work on collisionless shock waves debated not only about whether such phenomena could exist, but if they did exist, how such phenomena could form and evolve \citep[e.g., see][for detailed review]{wilsoniii14a, wilsoniii14b}.  The primary issue was that without collisions (i.e., a mechanism to dissipate energy, thus, produce entropy), a steepening sound wave should break rather than form into a stable discontinuity.  Similar to the radiation belts, early observations had limited resolution which initially led to the conclusion that small-scale, high frequency waves were not the dominant energy dissipation mechanism \citep[e.g.,][]{scudder86c}.  However, recent observations of large amplitude electrostatic \citep[e.g.,][]{breneman13a, wilsoniii07a, wilsoniii10a} and electromagnetic waves \citep[e.g.,][]{hull12a, wilsoniii12c, wilsoniii13a, wilsoniii14a, wilsoniii14b} at collisionless shocks have renewed interest in possible energy dissipation mechanisms for these nonlinear phenomena.

\indent  For instance, using high resolution observations \citet[][]{wilsoniii07a, wilsoniii14b} both found that the quasi-linear estimate for wave-particle (anomalous) collisions exceeded Coulomb collision rates by over 7--12 orders of magnitude.  Furthermore, recent simulations \citep[e.g.,][]{petkaki08a} have shown that the quasi-linear estimates themselves are 2--3 orders of magnitude too small.  The next question was, what is the source for such large amplitude waves?

\indent  Initial results by \citet[][]{wilsoniii07a} and \citet[][]{breneman13a} suggested a dependence of both occurrence and amplitude of electrostatic waves on changes in the magnetic field interpreted as currents.  A more recent two-part study \citep[i.e.,][]{wilsoniii14a, wilsoniii14b} found a direct relationship between the magnitude of the quasi-static current density, $\left\lvert \textbf{j}{\scriptstyle_{o}} \right\rvert$, and the wave electric field magnitude, $\left\lvert \delta \textbf{E} \right\rvert$ (results illustrated in Figure \ref{fig:exampledWWEvsJoExtrap}B).  Additionally, they found that current-driven electromagnetic emissions were not limited to just the shock ramp, which begs the question of the level of fundamental importance of such a relationship.

%%****************************************************************************************
%%  Subsection:  Currents and Waves
%%****************************************************************************************
\subsection{Currents and Waves}  \label{subsec:CurrentsandWaves}
\indent  The dependence of wave amplitude on the current density (in the terrestrial bow shock) is shown in Figure \ref{fig:exampledWWEvsJoExtrap}B.  The plot shows nearly $\sim 10^{6}$ points as a scatter plot comparing $\left\lvert \delta \textbf{E} \right\rvert$ with $\left\lvert \textbf{j}{\scriptstyle_{o}} \right\rvert$.  Color-coded contours have been over-plotted to differentiate regions with the highest density of points.  The color-coded legend in the lower-left hand corner corresponds to the color-coded points in the plot (e.g., points within outermost contour comprise $\sim$90\% of the data, i.e., the red$+$orange$+$purple points).  Three fit lines have been over-plotted, where the color of the line corresponds to the fraction of points used in the fit associated with the color-coded legend in the lower-left hand corner.  For instance, the red line corresponds to a fit of all points within the purple contour (i.e., the red$+$orange$+$purple points).  The fit parameters and fit function (i.e., $\delta E$ $=$ $A \ j{\scriptstyle_{o}}$ $+$ $\delta E{\scriptstyle_{o}}$) are shown in Figure \ref{fig:exampledWWEvsJoExtrap}C, which will be discussed in the next section.

\indent  The results show a general trend of increasing $\left\lvert \delta \textbf{E} \right\rvert$ with increasing $\left\lvert \textbf{j}{\scriptstyle_{o}} \right\rvert$, consistent with theories \citep[e.g.,][]{sagdeev66, gary81c} predicting that $\left\lvert \delta \textbf{E} \right\rvert$ results from currents radiating electromagnetic waves, called current-driven instabilities.  This phenomena is not limited to collisionless shock waves.  It appears that currents (i.e., observed as time variations and/or gradients in the magnetic field) are sources of free energy regardless of their location, as seen in the solar wind \citep[e.g.,][]{malaspina13a}.  The radiation of high frequency waves is not limited to currents, as evidenced by recent observations within the terrestrial magnetosphere \citep[e.g.,][]{malaspina15a}, suggesting small-scale waves play important roles in energy dissipation at multiple types of boundary layers.  The trend shown in Figure \ref{fig:exampledWWEvsJoExtrap}B begs the question:  How far can one extend this relationship for predictive purposes?

%%****************************************************************************************
%%  Subsection:  Extrapolation to Inaccessible Regions
%%****************************************************************************************
\subsection{Extrapolation to Inaccessible Regions}  \label{subsec:ExtrapolationtoInaccessibleRegions}
\indent  The trend found by \citet[][]{wilsoniii14b}, shown in Figure \ref{fig:exampledWWEvsJoExtrap}B, approximately satisfies a power-law relationship given by $\left\lvert \delta \textbf{E} \right\rvert$ $\propto$ $\left\lvert \textbf{j}{\scriptstyle_{o}} \right\rvert^{\gamma}$.  Here we will use this trend to predict ranges for $\left\lvert \delta \textbf{E} \right\rvert$ (e.g., results shown in Figure \ref{fig:exampledWWEvsJoExtrap}C) to currently inaccessible regions of space for the purpose of prediction and motivation of further research.

\indent  There are several candidates including the solar corona, neutron star magnetospheres, and the extremes of a magnetar.  Theoretical work estimates that the current densities in each of these regions are:  $10^{4}$--$10^{7}$ $\mu A \ m^{-2}$ for the solar corona \citep[e.g.,][]{regnier09a}; $10^{14}$--$10^{15}$ $\mu A \ m^{-2}$ for neutron stars \citep[e.g.,][]{goldreich69a}; and $10^{19}$--$10^{22}$ $\mu A \ m^{-2}$ for magnetars \citep[e.g.,][]{stepanov13a}.  We use these ranges of $\left\lvert \textbf{j}{\scriptstyle_{o}} \right\rvert$, the power-law relationship described above, and the fit parameters associated with the color-coded lines in Figure \ref{fig:exampledWWEvsJoExtrap}B to estimate ranges of $\left\lvert \delta \textbf{E} \right\rvert$.  The color-coded lines in Figure \ref{fig:exampledWWEvsJoExtrap}C correspond to the same color-coded fit lines in Figure \ref{fig:exampledWWEvsJoExtrap}B.  The average values are shown as squares and the error bars correspond to the ranges for $\left\lvert \textbf{j}{\scriptstyle_{o}} \right\rvert$ and $\left\lvert \delta \textbf{E} \right\rvert$.  While the neutron star and magnetar environments involve pair versus proton-electron plasmas, there should still be significant currents and associated radiated instabilities.  The point is to illustrate the potential for truly dramatic electric fields and to encourage questions about potential acceleration limits/possibilities.

\indent  The numerical ranges for $\left\lvert \delta \textbf{E} \right\rvert$ are shown to the right of the panel and color-coded for each fit line.  As a benchmark for validation, we find our estimated solar corona fields of $\sim$ 0.5--300 V/m are consistent with previous predictions \citep[e.g.,][]{davis77a}.  This provides confidence in our extrapolations to the next two more extreme environments.

\indent  Near neutron stars, we predict $\left\lvert \delta \textbf{E} \right\rvert$ ranging from $\sim$ $4 \times 10^{7}$ to $9 \times 10^{10}$ V/m.  The most striking result is that $\left\lvert \delta \textbf{E} \right\rvert$ near magnetars range from $\sim$ $1 \times 10^{10}$ to $6 \times 10^{15}$ V/m.  This is equivalent to saying that an electron falling through the equivalent electric potential for one meter, ignoring any radiative losses,  would gain $> 10^{15}$ eV of energy!  Thus, a particle could reach the so called ``knee'' of the cosmic ray spectrum \citep[e.g.,][]{horandel03a} within only one meter.

%%----------------------------------------------------------------------------------------
%%  Section:  Discussion
%%----------------------------------------------------------------------------------------
\section{Discussion}  \label{sec:Discussion}
\indent  The purpose of this commentary is to provoke thought and further investigation into the relative importance of electron-scale (e.g., few tens to hundreds of meters) wave-particle interactions in the dynamics of large-scale (e.g., few tens of kilometers or larger) systems and their potential for particle energization.  We have provided a few examples illustrating the potential for such waves to produce relativistic electrons in the radiation belts or even ultra-relativistic particles in the magnetospheres of magnetars.  We have also provided evidence that such waves can dissipate energy, thus affect the global dynamics of the macroscopic systems in which they exist.  We hope that the dramatic energies predicted in this commentary provoke a greater emphasis on the small-scale wave-particle interactions in future theoretical and observational studies. \\

%%----------------------------------------------------------------------------------------
%%  Acknowledgements
%%----------------------------------------------------------------------------------------
{\large \noindent \textbf{Acknowledgements}}  \\
\indent  We would like to thank A.F.- Vi{\~n}as, D. Bryant, D.A. Roberts, R.T. Wicks, R. Lysak, V. Krasnoselskikh, and M.L. Goldstein for useful discussions of the fundamental physics involved in our study.  The work was partially supported by \emph{Wind} MO\&DA grants and NASA ROSES H-GI grants.  The \emph{Wind} TDS data and software used in this paper are found in the level zero files in the University of Minnesota \emph{Wind} WAVES servers, available on individual request.  The THEMIS data used in this paper and the associated calibration software can be found at: \\ 
  \textcolor{Blue}{http://themis.ssl.berkeley.edu/index.shtml}; and \\
  \textcolor{Blue}{http://cdaweb.gsfc.nasa.gov}.

%%----------------------------------------------------------------------------------------
%%  Bibliography
%%----------------------------------------------------------------------------------------
\renewcommand{\bibsep}{2pt}	% Tighten spacing between references
%\bibliography{/Users/lbwilson/Desktop/Lynn_B_Wilson_III/LaTeX/Bibliographies/my_bib_maker}

\end{document}